\documentclass[aps, amsmath,amssymb,prx,nofootinbib]{revtex4-2}

\usepackage{dcolumn}
\usepackage{bm}
\usepackage{comment}
\usepackage{multirow}
\usepackage{tikz,graphics,color,fullpage,float,epsf,
}

\usepackage[caption=false]{subfig}
\usepackage{tablefootnote}

\usepackage{orcidlink}

\def\CCC{C$^{3}$~}

\def\ee{e^+e^-}

\def\CCCnospace{C$^{3}$}

\def\gwp{CO$_2$e~}
\def\gwpnospace{CO$_2$e}

\begin{document}


\title{Sustainability Strategy for the Cool Copper Collider}
\author{Martin Breidenbach\,\orcidlink{0000-0001-5022-0835}, Brendon Bullard\,\orcidlink{0000-0001-7148-6536}, Emilio Alessandro Nanni\,\orcidlink{0000-0002-1900-0778}, Dimitrios Ntounis\,\orcidlink{0009-0008-1063-5620}, and Caterina Vernieri\,\orcidlink{0000-0002-0235-1053}}

\affiliation{SLAC National Accelerator Laboratory, 2575 Sand Hill Road, Menlo Park, California 94025, USA \& \\ Stanford University, 450 Jane Stanford Way, Stanford, California 94305, USA}


\begin{abstract}
The particle physics community has agreed that an electron-positron collider is the next step for continued progress in this field, giving a unique opportunity for a detailed study of the Higgs boson.  Several proposals are currently under evaluation of the international community.  Any large particle accelerator will be an energy consumer and so, today, we must be concerned about its impact on the environment. This paper evaluates the carbon impact of the construction and operations of one of these Higgs factory proposals, the Cool Copper Collider.  It introduces several strategies to lower the carbon impact of the accelerator.  It proposes a metric to compare the carbon costs of Higgs factories, balancing physics reach, energy needs, and carbon footprint for both construction and operations, and compares the various Higgs factory proposals within this framework. For the Cool Copper Collider, the compact 8 km footprint and the possibility for cut-and-cover construction greatly reduce the dominant contribution from embodied carbon.
\end{abstract}

\maketitle{}

\section{Introduction}

An electron-positron collider gives a unique opportunity to study the Higgs boson's properties with unprecedented precision and also provide an exceptionally clean environment to search for subtle new physics effects~\cite{narain2022future}.  A number of different  "Higgs factory" (HF) proposals, based on linear and circular colliders, are now under consideration.  All of these provide $\ee$ collisions at center-of-mass energies ($\sqrt{s}$) in the range of 240-370 GeV, and some also are capable of reaching higher energies.  

A high-energy particle collider is a large energy-consuming research facility.  As such, it is important to balance its scientific importance against its environmental cost.  The environmental impact of large accelerators has been analyzed in the recent Snowmass 2021 study~\cite{Butler:2023eah} of the future of particle physics in the USA~\cite{Roser:2022mae,Bloom:2022wwv,Bloom:2022gux}.  References~\cite{Bloom:2022wwv,green_ilc,Janot:2022jtn,arup} have examined the environmental cost of particular Higgs factory proposals, although often concentrating on particular elements of the total cost.

In this paper, we attempt a comprehensive evaluation of the carbon cost of the Cool Copper Collider (\CCC) Higgs factory proposal ~\cite{C3,c3phyiscswhitepaper} over its full lifetime, including costs from construction and from operation over the proposed timeline.  This paper is structured as follows: in Sec.~\ref{sec:cccdesign}, we briefly review the design of \CCC.  In Section~\ref{sec:physics}, we review the physics reach for \CCC and other Higgs factory proposals and introduce a metric for balancing carbon impact against the physics impact of each proposal.  In Sec.~\ref{sec:costop}, we analyze the power costs of operation of \CCC and describe methods for modifying the power design of the accelerator that would lead to substantial savings with little impact on the physics performance.  In Sec.~\ref{sec:co2}, we analyze the carbon impact of the construction of \CCC and emphasize that cut-and-cover construction, as opposed to construction in a deep tunnel, has significant advantages.  In Sec.~\ref{sec:cost}, we discuss options for the source of electrical power for the \CCC laboratory.  In Sec.~\ref{sec:results},  we bring these analyses together to estimate the total carbon footprint of \CCC.  Using information from available studies and design reports, we estimate the carbon impact of other Higgs factory proposals and compare these to \CCC in the framework described in Sec.~\ref{sec:physics}. 

\section{Review of the \CCC accelerator design}
\label{sec:cccdesign}

\CCC, which was proposed recently~\cite{C3,c3phyiscswhitepaper}, is a linear facility that will first operate at 250 GeV center-of-mass collisions. Immediately afterwards, without further extension of the linac, it will run at 550 GeV with an rf power upgrade. High energy operations will enable the exploration of the Higgs boson-top quark coupling, and provide direct access to the Higgs boson self-coupling with double Higgs boson production~\cite{c3phyiscswhitepaper,nanni2022c}. Furthermore, the beam polarization, which exploits the strong dependence of electroweak processes on the chirality of the initial state particles, will offer unique insights into the underlying physics, acting as a new tool for discovery~\cite{ILCInternationalDevelopmentTeam:2022izu}. This offers \CCC a strong complementarity with proton and circular $\ee$ colliders, where beam polarization is not possible. 

For \CCC, an approach radically different from the one adopted for linacs is used to build a collider with high gradient and high rf efficiency, and thus lower capital and operating costs~\cite{bane2018advanced}. \CCC is based on a distributed coupling accelerator concept, running under liquid nitrogen~\cite{nasr2021experimental}, that has led to an optimized accelerating gradient and minimized breakdown problems with respect to earlier designs based on normal conducting technologies. This has yielded an overall optimization of the gradient at 70 and 120 MeV/m for the 250 and 550 GeV operating points, respectively\cite{Belomestnykh:2022hlx}. Much higher energies are possible if length is not the major consideration. The fundamental \CCC parameters, assumed for the analysis in this paper, are shown in Table~\ref{tab:LCparam}. 

By far the major development to date is the actual distributed coupling accelerator structure. \CCC will use C-band (5.712 GHz) standing wave rf accelerating structures that are 1 m long. Each has an rf waveguide to bring power in, and in the more probable operating modes, splits rf power evenly between the beam and dissipation in the structure with 43\% beam loading.  Operating at 80 K brings the shunt impedance up to 300 M$\Omega$/m, allowing for efficient operation at 120~MeV/m.  These gradients have been demonstrated at C-band~\cite{schneider2022high} and with an electron beam in an X-Band (11.424 GHz) structure on the SLAC XTA beamline~\cite{nasr2021experimental}. The C-band structure has been tested at low power at SLAC and at high power without beam at Radiabeam~\cite{RBoct}. The \CCC gradient results in a collider with a 550 GeV center-of-mass energy capability on an 8 km footprint. 

A preconceptual design for the overall linac cryogenic system has been developed that includes the design for the cryomodules. For the \CCC 250 and 550 GeV design, each linac will have three re-liquification cryoplants. Liquid nitrogen will flow out along the linac in both directions, so there are six flow runs. The liquid nitrogen will be above the raft structures, with an initial velocity of $\sim$0.03 m/s. The liquid nitrogen will cool the accelerator structures by nucleate boiling with a power density of 0.4 W/cm$^2$, producing saturated vapor which counter-flows back to the cryoplant.  Each cryo-run is about 450 meters in length. The vapor velocity near the cryoplant is $\sim$3 m/s.

\begin{table}[ht!]
\caption{Target beam parameters for \CCCnospace. } 
 \label{tab:LCparam}
\begin{center}
\begin{tabular}{c  c  c } 
\hline\hline
 Parameter [Unit] & \multicolumn{2}{c}{Value} \\
 \hline
   $\sqrt{s}$ (GeV) & 250 & 550 \\
   Luminosity (cm$^{-2}$ sec$^{-1}$) & $1.3 \cdot 10^{34}$ &  $2.4 \cdot 10^{34}$  \\
  Number of bunches per train  & 133-200  & 75 \\
  Train repetition rate (Hz) &  120 & 120 \\
  Bunch spacing (ns) &  5.3-3.5\footnote{Beam dynamics and structure optimization studies are ongoing: the injected charge range is between 0.7 and 1 nC with a bunch spacing ranging between 3.5 and 5.3~ns and final horizontal beam size ranging between 156.7~nm and 182~nm, without changes to the instantaneous luminosity. Accelerating structure optimization studies include varying the $a/\lambda$ from 0.05 to 0.07  with $\pi$-2$\pi$/3 phase advance~\cite{bane2018advanced} to reduce the longitudinal wakefield and preserve the shunt impedance.} 
  &  3.5  \\
  Site power (MW)  & 150 & 175  \\
  Beam power (MW) & 2.1 & 2.45\\
Gradient (MeV/m) & 70 & 120 \\
Geometric Gradient (MeV/m) &  63 & 108 \\   %
rf Pulse Length (ns) & 700 & 250 \\
Shunt Impedance (M$\Omega$/m) &  300 & 300 \\ %
Length (km)  & 8 & 8 \\
\hline\hline
 \end{tabular}
\end{center}
\end{table}

\section{Comparison of Higgs factory physics reach} \label{sec:physics_reach}
\label{sec:physics}
Among the $\ee$ colliders being evaluated by the community, the International Linear Collider (ILC)~\cite{ILC, ILCInternationalDevelopmentTeam:2022izu}, based on superconducting rf technology, has the most advanced design~\cite{Roser:2022sht}, and the ILC is currently under consideration for construction in Japan. 
CERN is pursuing as its main strategy  a large circular collider, the Future Circular Collider (FCC)~\cite{fcc_snowmass}, and China is planning a similar circular collider, the Circular Electron Positron Collider (CEPC)~\cite{Gao:2022lew}.  Each of these circular colliders would require a tunnel with circumference of the order of 100 km to limit synchrotron radiation.  However, the expected instantaneous luminosity drops off significantly above center-of-mass energies of 350--400 GeV.  
An alternative strategy is to construct a compact linear $\ee$ collider based on high gradient acceleration. CERN is also pursuing such a proposal, the Compact Linear Collider (CLIC)~\cite{clic}, that would operate at a collision energy of 380 GeV.  

The carbon footprint of the proposed future Higgs factories should be assessed relative to the expected physics reach, which has been reviewed most recently in the context of the Snowmass Process~\cite{narain2022future,dawson2022report}. The primary physics goal of a future Higgs factory is the determination of the total Higgs width and Higgs couplings with per-cent or sub-per-cent precision. A reasonable figure of merit to gauge the physics reach of each machine is the expected level of precision for each of these measurements. We note that evaluating the projected measurement precision accounts for the fact that different beam configurations (center-of-mass energy and beam polarization) have a strong impact on the physics reach of each of those machines. These differences in precision are not accounted for when comparing the total number of Higgs bosons produced alone~\cite{narain2022future,ILCInternationalDevelopmentTeam:2022izu}. 

The physics reach at $\ee$ colliders increases with the center-of-mass energy, since different Higgs boson production mechanisms become accessible. At 250 GeV center-of-mass energy operations the main Higgs boson production mechanism is associated production with a Z boson ($\ee \rightarrow ZH$), enabling a model-independent determination of the Higgs boson total width. Higgs boson production via the $W$-boson fusion reaction $e^{+}e^{-} \rightarrow \nu \bar{\nu}H$ is accessible at $\sqrt{s}\sim 500$ GeV, where the only visible signals in the final state come from Higgs boson decays. This allows Higgs boson measurements governed by different systematic effects, complementary to the $250 \ \mathrm{GeV}$ data, as well as opportunities to study effects such as separation of $H \rightarrow gg/b\bar{b}/c\bar{c}$ decays and CP violation in $H \rightarrow \tau^{+}\tau^{-}$ \cite{ILCInternationalDevelopmentTeam:2022izu}. Importantly, at high center-of-mass energies, double Higgs boson production in the $ZHH$ channel opens up, providing direct access to the Higgs boson self-coupling $\lambda_{3}$. Direct constraints on the top quark Yukawa coupling $y_t$ also become accessible via $e^+e^-\rightarrow t\bar{t}H$ production at $\sqrt{s}$ above approximately 500 GeV. At circular machines, given the energy limitations, double Higgs boson and associated top production mechanisms are not accessible, thus allowing only for indirect and model-dependent measurements of $\lambda_{3}$ and $y_t$, both through loop effects in single Higgs boson production.

The use of longitudinal beam polarization offers unique advantages for effective precision measurements at a linear $\ee$ collider, since the interaction cross sections at an $\ee$ collider have strong dependencies on beam polarization. 
It has been demonstrated that at 250 GeV center-of-mass energy, the ultimate precision reach in the determination of Higgs couplings, through a Standard Model Effective Field Theory (SMEFT) analysis, for an integrated luminosity of $2 \ \mathrm{ab}^{-1}$ with polarized beams, is comparable to that of $5 \ \mathrm{ab}^{-1}$ with unpolarized beams, with most of the gain coming from $e^{-}$ polarization alone~\cite{Barklow:2017suo}.   The main effect of beam polarization is to discriminate the effect of different SMEFT operators that contribute to the Higgs boson coupling.  There is a similar gain of about a factor of 2.5 from discrimination of the effects of the operators contributing to the $WW\gamma$ and $WWZ$ couplings, which also enter the SMEFT analysis.
The positron polarization becomes more relevant at higher center-of-mass energies. For instance, W-boson fusion reactions, such as $e^{+}e^{-} \rightarrow \nu \bar{\nu}H$, proceed only from $e_{L}^{-}e_{R}^{+}$ initial states, providing a cross-section (or, equivalently, effective luminosity) enhancement of $\sim 2.5$ for typical polarizations foreseen at future linear $\ee$ machines \cite{ILCInternationalDevelopmentTeam:2022izu}.  Here positron polarization makes a significant contribution. This implies that the same number of Higgs bosons can be produced through this process with only  $\sim 40 \%$ of the integrated luminosity, compared to having unpolarized beams. 

Direct measurement of CP-odd observables would unambiguously establish CP-violation in the top-Higgs coupling. These are challenging at the Large Hadron Collider (LHC) due to the inability to fully reconstruct the $t\bar{t}H$ system.
The use of polarized beams has been shown to boost the sensitivity to the CP structure of the top quark-Higgs boson coupling in combined measurements of total cross section of $\ee\rightarrow t\bar{t}H$ production~\cite{Gadbole2011} and related CP-violating observables. These direct probes are complementary to model-dependent constraints on CP-mixed fermion-Higgs couplings from measurements of electric dipole moments~\cite{Bahl2022}.  

Moreover, beam polarization at high energy enables the suppression of relevant backgrounds, such as the dominant $e^{+}e^{-} \rightarrow W^{+}W^{-}$ background for positive (negative) electron (positron) beam polarization, increasing the signal-over-background ratio and allowing the precise measurement of the rate of other backgrounds, as well as the reduction of detector-related systematic uncertainties, with combined measurements of datasets with four distinct initial-state polarization configurations. These effects collectively indicate the increased precision reach that beam polarization provides for linear machines~\cite{ILCInternationalDevelopmentTeam:2022izu}.

For these reasons, in this analysis we propose a comparison of the carbon footprint of collider concepts relative to their expected precision in Higgs coupling measurements. Table~\ref{tab:higgs_couplings} summarizes the projected relative precision for Higgs boson couplings measurements at each collider combined with projected results from the High Luminosity Large Hadron Collider (HL-LHC). As can be seen, the overall physics reach of all proposed Higgs factories is similar~\cite{dawson2022report,narain2022future} for operations at 240-250 GeV, and additional measurements become accessible for the higher center-of-mass energy runs at linear colliders. We also compare the Higgs Factory proposals is in terms of total energy consumption and carbon emissions, for both construction activities and operations, which is the most relevant quantity when one is evaluating each project's impact on the global climate.

\begin{table}[!ht]
    \centering
       \caption{Relative precision (\%) of Higgs boson coupling and total Higgs boson width measurements at future colliders when combined with HL-LHC measurements. Results are from Ref.~\cite{dawson2022report}. The FCC-ee numbers assume two  interaction points (IPs) and 5 ab$^{-1}$ at 240 GeV and 1.5 ab$^{-1}$ at 365 GeV.  The CEPC numbers also assume two IPs, but 20 ab$^{-1}$ at 240 GeV and 1 ab$^{-1}$ at 360 GeV. The top quark Yukawa coupling can be measured with nearly double the precision at \CCC operations at 550 GeV, compared to the ILC operating at 500 GeV, because of the higher center-of-mass energy~\cite{Barklow:2015tja}. Nevertheless, in this study we assume the same precision for \CCC as for the ILC at 500 GeV. Note that since there are no beyond the Standard Model decays allowed in this table, the width is constrained by the sum of the Standard Model contributions. Entries with a dash (-) correspond to couplings that are out of reach ($hc\bar{c}$ at HL-LHC) or for which estimates were not yet available at the time of writing ($hhh$ for CEPC). The weighted average shown in the last row has been calculated as explained in the main text. }
    \label{tab:higgs_couplings}
     \resizebox{\columnwidth}{!}{%
    \begin{tabular}{c|c|c|c|c|c|c}
    \hline \hline 
    &  & \multicolumn{5}{c}{HL-LHC + } \\ \hline
    Relative Precision $(\%)$ & HL-LHC &  CLIC-380 & ILC-250/C$^3$-250 & ILC-500/C$^3$-550  & FCC 240/360  & CEPC-240/360  \\ \hline
\hline$h Z Z$ & 1.5 & 0.34 & 0.22 & 0.17 & 0.17 &   0.072  \\
$h W W$ & 1.7 & 0.62 & 0.98 & 0.20 &  0.41 &  0.41   \\
$h b \bar{b}$ & 3.7 & 0.98 & 1.06 & 0.50 &  0.64 & 0.44   \\
$h \tau^{+} \tau^{-}$ & 3.4  & 1.26 & 1.03 & 0.58 &  0.66 &  0.49   \\
$hgg$ & 2.5 & 1.36  & 1.32 & 0.82  & 0.89 & 0.61  \\
$h c \bar{c}$ & - & 3.95 & 1.95 & 1.22  & 1.3 & 1.1   \\
$h \gamma \gamma$ & 1.8 & 1.37 & 1.36 & 1.22 & 1.3 & 1.5   \\
$h \gamma Z$ & 9.8 & 10.26 & 10.2 & 10.2 &  10  & 4.17   \\
$h \mu^{+} \mu^{-}$ & 4.3  & 4.36 & 4.14 & 3.9 &  3.9  & 3.2   \\
$h t \bar{t}$ & 3.4 & 3.14 & 3.12 & 2.82/1.41  & 3.1 &  3.1   \\
$hhh$ & 50 &  50 & 49 &  20 & 33 & -   \\
$\Gamma_{\mathrm{tot}}$ & 5.3 & 1.44  & 1.8 & 0.63 & 1.1 &  1.1  \\
\hline \hline
Weighted average & - & 0.94 & 0.86 & 0.45 & 0.59 & 0.49 \\
\hline \hline
\end{tabular}   
    }
\end{table}

We then present an estimate of energy consumption and carbon footprint per unit of physics output. This is achieved by taking the average of the relative precision over all Higgs couplings, weighing them by the relative improvement in their measurement with respect to HL-LHC:

\begin{equation}
    \left \langle \frac{\delta \kappa}{\kappa} \right \rangle = \frac{\sum\limits_{i}{w_{i}\left(\frac{\delta \kappa}{\kappa}\right)_{i}}}{\sum\limits_{i}{w_{i}}}
    \label{eq:weighted_precision}
\end{equation}

where the sum runs over the entries in each column of Table~\ref{tab:higgs_couplings} and the weight is defined as:

\begin{equation}\label{eq:weight}
    w = \frac{\left(\frac{\delta \kappa}{\kappa}\right)_{\mathrm{HL-LHC}}-\left(\frac{\delta \kappa}{\kappa}\right)_{\mathrm{HL-LHC}+\mathrm{HF}}}{\left(\frac{\delta \kappa}{\kappa}\right)_{\mathrm{HL-LHC}+\mathrm{HF}}}
\end{equation}

This definition weights measurements by their relative improvement over the HL-LHC measurements when the HL-LHC and  future Higgs Factory results are combined. Qualitatively, measurements that minimally improve those of HL-LHC are assigned weights near zero, while HF measurements with high precision or large improvement over HL-LHC are assigned larger weights. While other weighting schemes could be used, we argue that Eq.~\ref{eq:weight} is unbiased towards the type of physics measurement (e.g. Yukawa, self-coupling, vector coupling) and it emphasises the individual strengths of each collider facility. 

For the estimation of the weighted average precision, the $hc\bar{c}$ coupling was excluded, since there is no estimate for HL-LHC, whereas we assume that the $hhh$ coupling for the CEPC can be measured with the same precision as for the FCC. The weighted average precision for each collider is given in the last row of Table~\ref{tab:higgs_couplings}.

\section{Power consumption and optimizations}
\label{sec:costop}

The most obvious way to reduce the carbon impact of a major facility is to minimize the amount of power that it consumes, thereby minimizing the associated emissions from energy production. This is firmly within the means of the facility designers and crucially does not rely on grid electrification. The nominal operating parameters for \CCC operating at 250 GeV are given in Table~\ref{tab:powerparam}.  

\begin{table}[ht!]
\caption{Estimated electrical power requirements for \CCC operating at 250 GeV at nominal luminosity.} 
 \label{tab:powerparam}
\begin{center}
\begin{tabular}{l  r } 
\hline\hline
 Parameter (Unit) & Value \\
 \hline
   Temperature (K) & 80 \\
   Beam Loading (\%)  & 45 \\
   Gradient (MeV/m) & 70  \\
   Flat Top Pulse Length ($\mu$s) & 0.7 \\
  
   \hline
    Total Beam Power (MW) & 4 \\ 
     Total rf Power (MW) & 18 \\
     Heat Load at Cryogenic Temperature (MW) & 9 \\
     Electrical Power for rf (MW) & 40 \\
     Electrical Power For Cryo-Cooler (MW) & 60 \\
     Accelerator Complex Power (MW) & $\approx$ 50 \\
   
 \hline
  Site Power (MW) & $\approx$ 150 \\
  \hline\hline
 \end{tabular}
\end{center}
\end{table}


Several avenues can be pursued to optimize operational power requirements. Increases in luminosity or reduction in power consumption are possible through the development of ancillary technology by increasing the rf source efficiency, increasing the efficiency of powering the accelerating structures or modification of beam parameters to increase luminosity. At present the main linac requires $\sim$100~MW of power with 40~MW for the rf sources and 60~MW for the cryogenic system.

For the rf sources, the \CCC concept utilizes an overall rf system efficiency of 50\% which is in line with present high power rf sources that are designed with efficiency in mind. However, significant advances in modern design techniques for klystrons are increasing the klystron amplifier's ultimate efficiency significantly with the inclusion of higher-order-mode cavities, multicell outputs and advanced multidimensional computational tools. For example, designs now exist for a 50~MW class rf source~\cite{cai2020design} approaching an amplifier efficiency of 70\%. Multibeam rf sources, reducing the beam perveance, have rf source designs exceeding 80\% efficiency~\cite{chao2022modeling}. These results reinforce modern understanding on the limits of klystron efficiency~\cite{cai2021XBandKly} which indicate a klystron amplifier efficiency of 70-80\% is possible, leading to an overall rf source efficiency of 65\%.

Radio-frequency pulse compression, presently not in the \CCC baseline, is also a well-known technique for powering high gradient structures. For \CCCnospace, pulse compression is particularly useful due to the impact of power loss at cryogenic temperatures and due to the relatively long fill time for a copper structure operating at cryogenic temperatures. In a previous study~\cite{bane2018advanced}, it was found that low factors of pulse compression, which preserves rf efficiency in the compressor~\cite{wang2017rf}, improves the overall efficiency of the system by 30\%. Recently, additional efforts have been made to realize the extremely high Q cavities required for pulse compression with cryogenically cooled rf structures~ \cite{snively2022cryogenic,schneider14th}; these include concepts operating at room temperature and inside the cryostat at 80~K.

\begin{figure}
    \centering
    \includegraphics[width=0.5\textwidth]{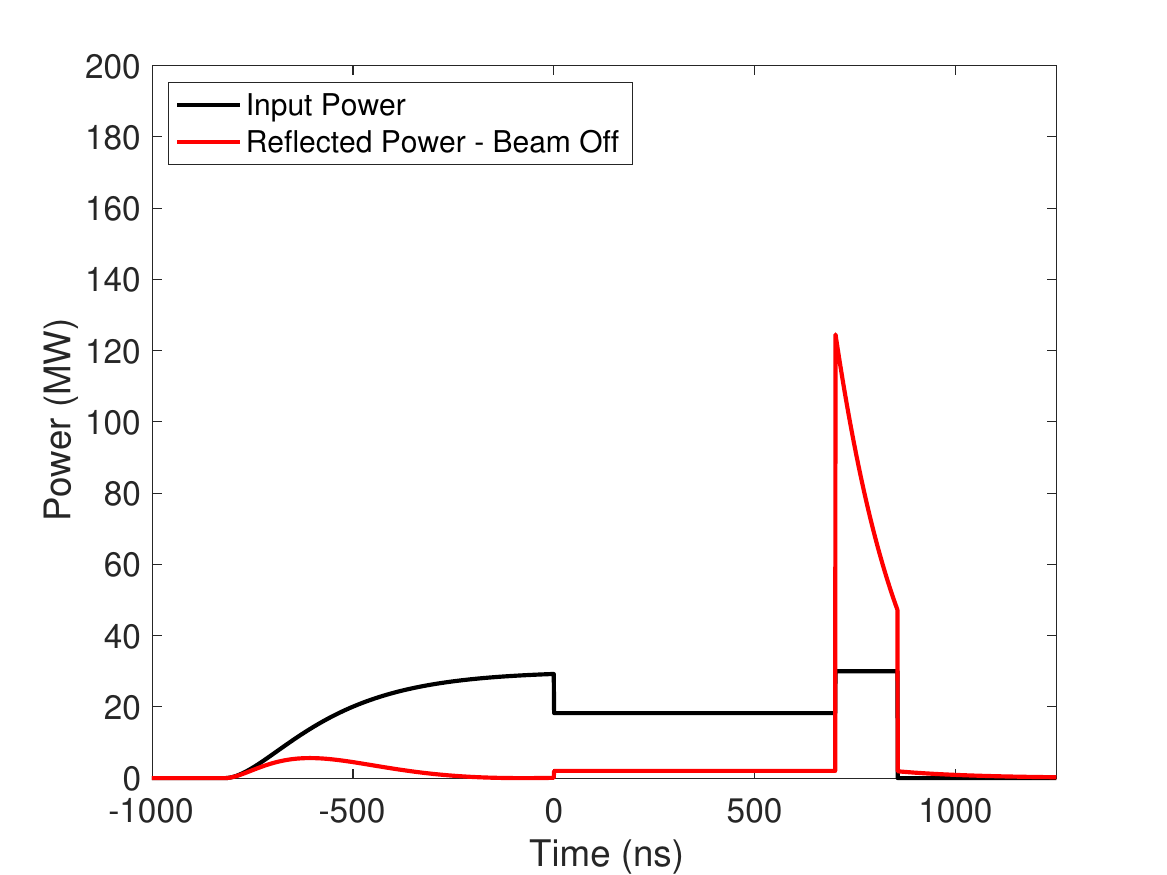}
    \caption{Forward and reflected power for 1 m of structure for operation at 70~MeV/m. An rf pulse is shown in the absence of the beam. With the beam the flat top power is constant at 30~MW.}
    \label{fig:powerpulse}
\end{figure}

\begin{figure}
    \centering
    \includegraphics[width=01\textwidth]{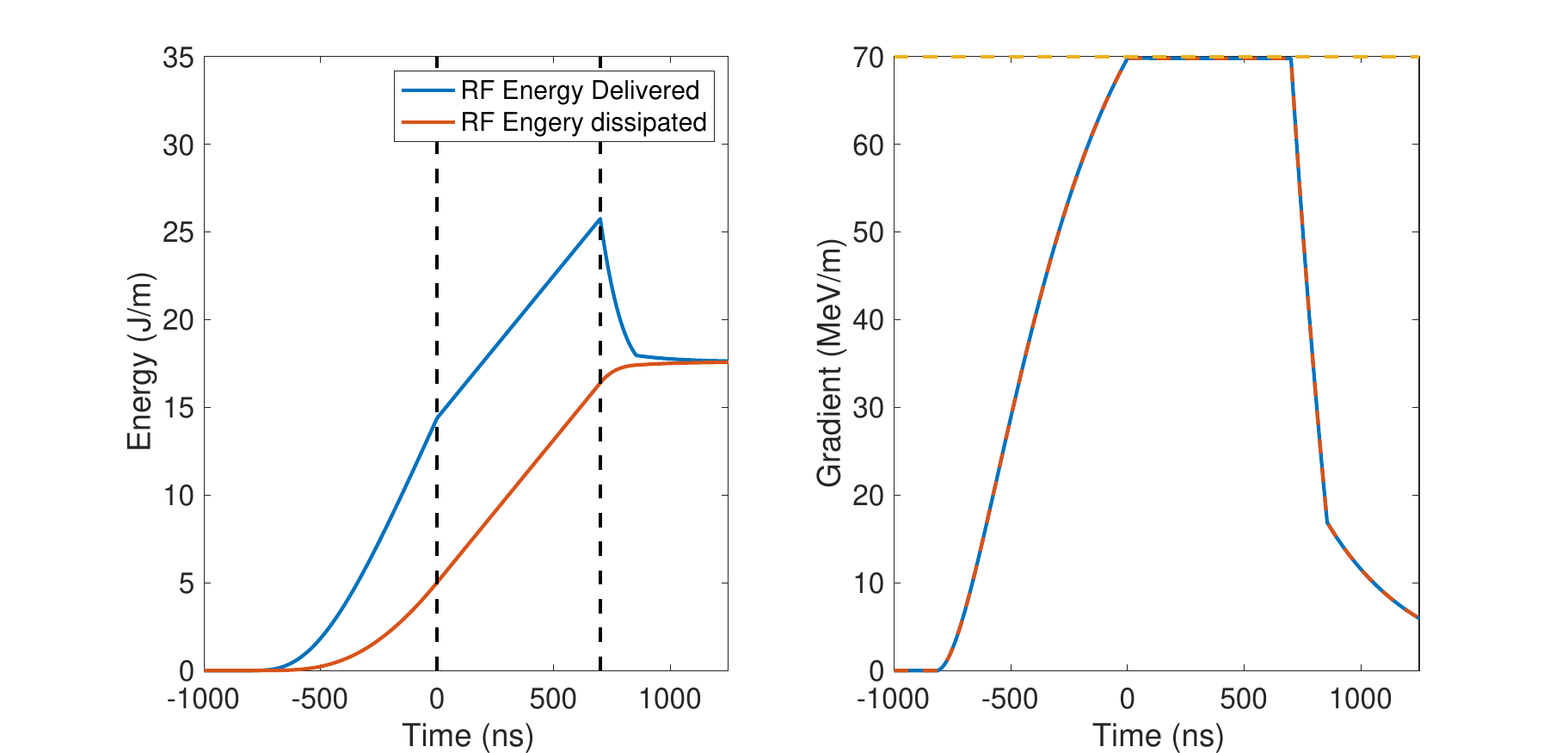}
    \caption{Dissipated energy per pulse per meter and the time domain gradient in the structure for a 70~MeV/m flat top for 700~ns.}
    \label{fig:eneregypulse}
\end{figure}

For the baseline \CCC design~\cite{C3,c3phyiscswhitepaper} we anticipate operation with 700 and 250~ns flat tops respectively for gradients of 70 and 120 MeV/m and a constant power dissipation of 2.5~kW/m at 120~Hz. Figure ~\ref{fig:powerpulse} and Figure~\ref{fig:eneregypulse} show the rf power, dissipated energy and gradient during the pulse. While these flat top lengths were selected to limit the challenges of breakdown, increasing the flat top length and reducing the repetition rate should be investigated in order to reduce the thermal load on the linac. At present, the thermal balance between the structure fill/dump time and the flat top is approximately 50\% (equal thermal load). If we were to extend the flat top lengths by a factor of two and reduce the repetition rate by a factor of two, the thermal dissipation in the main linac would decrease by 25\%. This improvement would have little effect on the overall design of the accelerator, and would be acceptable if the breakdown rates remain low enough. Proving that this is possible will require high gradient testing of structures with 1400 and 500~ns respectively.

The beam current of \CCC is relatively low thanks to the large bunch spacing and efficient accelerating structures. One could pursue the possibility of reducing the bunch spacing to increase the current. However, this will require compatibility studies with the detector design. Here we consider the scenario where the bunch spacing is reduced by a factor of two. This would keep a bunch spacing of more than 1~ns for \CCC operating at both 250 and 550 GeV, resulting in a decrease of 25\% for the cryogenics power. The rf power required would only decrease by 20\% because the peak rf power required would be slightly higher during the rf pulse flat top to compensate for the additional current.

We note that these approaches can all be combined for mutual benefit as shown in the last row of Table~\ref{table:power_savings}. The demonstration research and development plan~\cite{nanni2022c} will be able to investigate these approaches and lead to potential power savings.

\begin{table}[h!]
\caption{Power savings with adjustment in main linac design and beam parameters. For 550~GeV the percentage savings would be unchanged for a combined 79~MW reduction in electrical power from the nominal 125~MW for the main linac.}
  \label{table:power_savings}
  \centering
  \begin{tabular}{c c c c c}
    \hline\hline
   Scenario & rf System & Cryogenics & Total & Reduction \\
    \hline
   & (MW) & (MW) & (MW) & (MW) \\
    \hline
    Baseline 250 GeV  & 40 & 60 & 100 & - \\
    rf Source Efficiency Increased 15\%  & 31 & 60 & 91 & 9 \\
    rf Pulse Compression & 28 & 42 & 70 & 30 \\
    Double Flat Top  & 30 & 45 & 75 & 25 \\  
    Halve Bunch Spacing  & 34 & 45 & 79 & 21 \\  
    \hline\hline
     All Scenarios Combined  & 13 & 24 & 37 & 63 \\  
    \hline\hline
  \end{tabular}
\end{table}

The research and development needed to improve the operational efficiency of \CCC is recognized in the rf technology and beam physics community roadmaps for the U.S. Department of Energy~\cite{Bai:2022iem,Nagaitsev:2021rfs,Belomestnykh:2022hlx}.

\section{Carbon impact of construction}
\label{sec:co2}
Under the assumption that the electric grid will be successfully decarbonized by 2040, as it is the goal of many international climate plans, then construction, rather than operations, may well dominate the climate impact of a new particle physics facility~\cite{Bloom:2022gux}. 
For FCC it is projected that the whole accelerator complex~\footnote{The main tunnel plus the additional buildings on the site, the materials for the accelerator and detectors, assuming a main tunnel length of 97.7 km (the updated FCC design anticipates 91 km).} will have a carbon impact similar to that of the redevelopment of a neighbourhood of a major city~\cite{Bloom:2022gux}. This indicates that the environmental impact of any future collider facility is going to receive the same scrutiny as that of a major urban construction project. 
The bottom-up analysis in Ref.~\cite{Bloom:2022gux} derives an estimate of global warming potential (GWP) for the main tunnel material (concrete) manufacture alone to be equivalent to the release of 237 ktons of CO$_2$ (\gwpnospace). An alternative top-down analysis is instead dependent on the character of the earth to be excavated, leading to estimates ranging from 5-10 kton \gwpnospace/km of tunnel construction and total emissions of 489-978 kton \gwpnospace~\footnote{Contributions from many bypass tunnels, access shafts, large experimental caverns, and new surface sites are excluded.}. 

A life cycle assessment of the ILC and CLIC accelerator facilities is being performed by ARUP~\cite{arup} to evaluate their holistic GWP, so far providing a detailed environmental impact analysis of construction. The components of construction are divided into classes: raw material supply, material transport, material manufacture, material transport to work site, and construction process. These are labelled A1 through A5, where A1-A3 are grouped as materials emissions and A4-A5 are grouped as transport and construction process emissions. The total GWP for ILC and CLIC is 266 and 127 kton \gwpnospace~\cite{arup}, respectively\footnote{We use the emissions figures associated to the CLIC drive-beam design, which is more efficient than the alternative design utilizing only klystrons for rf power.}. The approximate construction GWP for the main tunnels are 6.38 kton \gwpnospace/km for CLIC (5.6m diameter) and 7.34 kton \gwpnospace/km for ILC (9.5m diameter); the FCC tunnel design is similar to that of CLIC, so 6.38 kton \gwpnospace/km is used for the calculation of emissions for both FCC and CEPC. While a comprehensive civil engineering report is unavailable for FCC and CEPC, we estimate the concrete required for klystron gallery, access shafts, alcoves, and caverns to contribute an additional 30\% of emissions, similar to what is anticipated for CLIC. The analysis indicates that the A4-A5 components constitute 20\% for CLIC and 15\% for ILC. In the absence of equivalent life cycle assessment analysis for FCC and CEPC, we account for the A4-A5 contributions as an additional 25\%. A summary of these parameters is given in Table~\ref{tab:const-emissions}.

\begin{table}[h]
    \caption{Summary of GWP for different collider proposals separated by origin. Note that FCC and CEPC emissions are estimated based on the comprehensive life cycle assessment of ILC and CLIC, whereas those of ILC and CLIC are directly quoted from the ARUP report~\cite{arup}. The assumptions for the \CCC estimate are discussed in the main text.}
    \label{tab:const-emissions}
    \centering
    \begin{tabular}{c | c | c  c  c }
    \hline\hline
    \multirow{2}{*}{Project} & \multirow{2}{*}{Main tunnel length (km)} & \multicolumn{3}{c}{GWP (kton \gwpnospace)} \\
    & & Main tunnel & + other structures & + A4-A5 \\
    \hline
    FCC & 90.6 & 578 & 751 & 939 \\
    CEPC & 100 & 638 & 829 & 1040 \\
    ILC & 13.3 & 97.6 & 227 & 266 \\
    CLIC & 11.5 & 73.4 & 98 & 127 \\
    \hline\hline
    \CCC & 8.0 &  133  & 133 & 146  \\
     \hline\hline
    \end{tabular}
\end{table}

\begin{figure}[h!]
    \centering
    \includegraphics[width=0.7\textwidth]{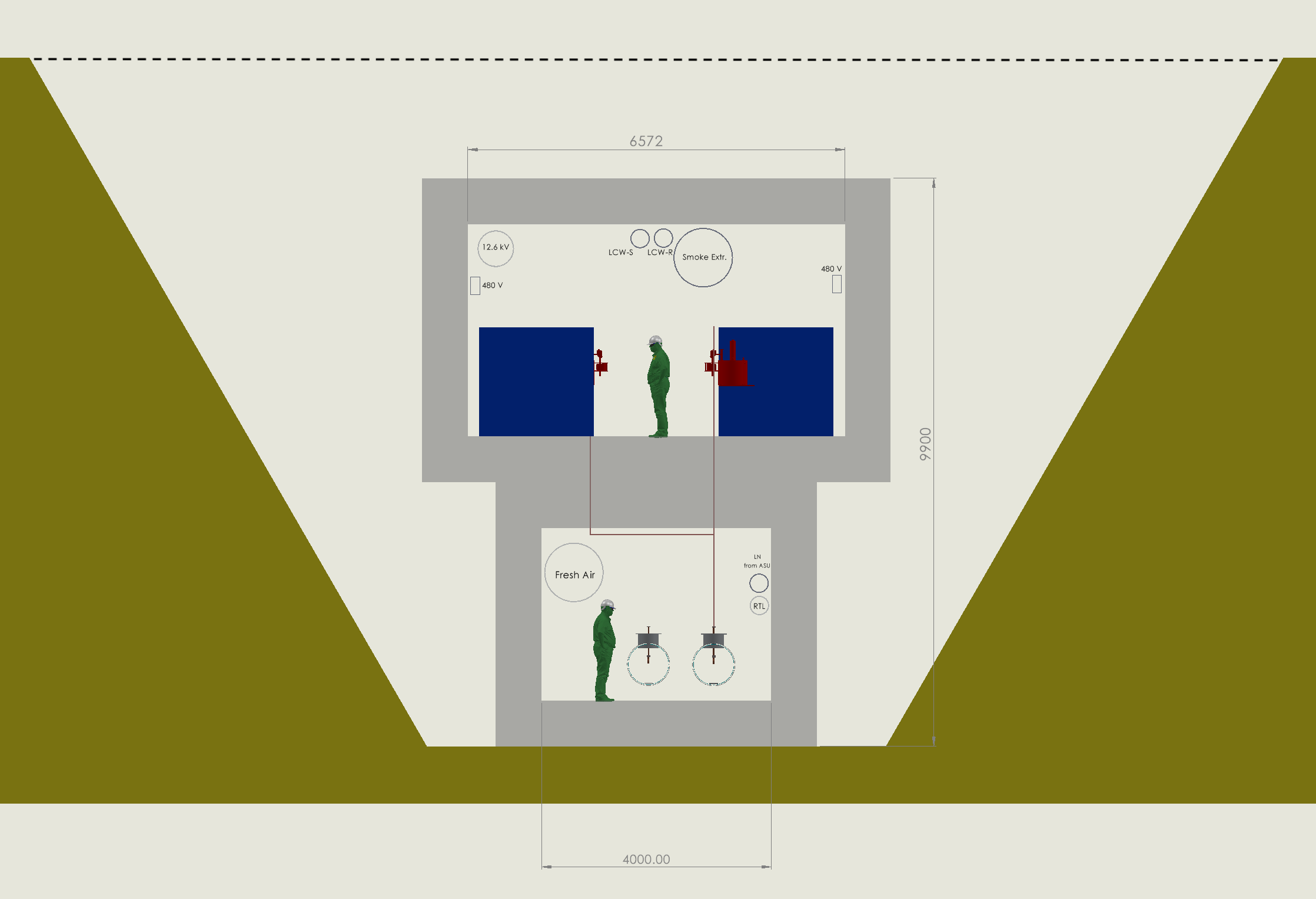}
    \caption{Layout of the \CCC klystron gallery (upper level) and accelerator hall (lower level) in the cut-and-cover construction approach, which is used for both the main linac and injectors. All dimensions are in mm. Key components of physical infrastructure are shown. The dashed line shows the ground level. All the excavated material will be placed to yield a small berm. Possible locations for Low Conductivity Water Supply (LCW-S),Low Conductivity  Water Return (LCW-R), Liquid Nitrogen makeup (LN) from the Air Separation Unit (ASU), and the Ring to Linac return line for the damped 10 GeV beam (RTL) are shown.  }
    \label{fig:cutcover}
\end{figure}

The \CCC tunnel will be about 8 km long with a rectangular profile in each of its component systems. Assuming a cut and cover approach, all the excavated material will be replaced to yield a small berm. We estimate that for the whole accelerator complex only about 50000 m$^3$ of spoil for the experimental hall will have to be relocated. Figure~\ref{fig:cutcover} shows a schematic of the \CCC cross section, where the klystron gallery is situated directly above the accelerator hall with sufficient concrete shielding to allow constant access to the klystron gallery during operation. The application of a top-down estimate of 6-7 kton \gwpnospace/km obtained from the ARUP report is not appropriate for the \CCC surface site due the differing cross section geometries of the accelerator housing. To allow for a fair comparison among facilities, we take the same basic assumptions of construction materials. In particular, that construction uses a mix of CEM1 C40 concrete and 80\% recycled steel, the GWP of concrete is taken to be 0.18 kg~\gwp/kg concrete with density 2400 kg/m$^3$~\cite{CEM1C40}, and 85\% of emissions originate from concrete and 15\% of emissions originate from steel production. Taking into account construction of the main linacs, injector linacs, damping rings, beam delivery system, and experimental hall, the total volume of construction material is estimated to be about 260000 m$^3$ (consisting mostly of concrete by volume). This leads to a GWP of 133 kton \gwp for A1-A3 components and GWP per unit length of the main linac of around 17 kton~\gwpnospace/km. Notably, this is roughly a factor two larger than the GWP/km of main tunnel construction of ILC and CLIC; this suggests further tunnel geometry optimizations are achievable with a detailed engineering study. The surface site construction eliminates the need for additional infrastructure (e.g. access tunnels and turnarounds) and greatly reduces the complexity of the construction process, which we estimate to account for an additional 10\%~\footnote{Nearly all spoil from the surface site excavation in the cut-and-cover approach is replaced onto the accelerator housing yielding a small berm. This removes the need to truck large volumes of spoil off-site, which would be done for the tunneled facilities ILC and CLIC. A surface site also eliminates the need to operate a boring machine. We therefore estimate the relative contribution of the A4-A5 component to \CCC construction emissions to be half that of the tunneled facilities.} to the GWP. This yields a final estimate of 146 kton~\gwp for civil engineering. 


Unlike other Higgs factories under evaluation, the \CCC site has not been decided yet. A \CCC $\ee$ collider could in principle be sited anywhere in the world. 
A community decision will be made regarding the actual site selection, although we note that the \CCC offers a unique opportunity to realize an affordable energy frontier facility in the USA in the near term and the entire program could be sited within the existing U.S. National Laboratories. The \CCC tunnel layout would be adapted to its location, and a cut and cover site, suitable for a horizontal layout, is extremely attractive also for both cost and schedule reasons.
The details of the siting options at Fermi National Accelerator Laboratory are discussed in~\cite{FCGwhitepaper}. Sites such as the U.S. Department of Energy Hanford site located in the Pacific Northwest have room to accommodate even bigger footprint machines within their site boundary.

\section{Mitigation strategies for operations} 
\label{sec:cost}

There can be considerable emissions associated with the production of energy required to meet site operation power requirements. This is highly dependent on the region in which the project operates; regions with highly decarbonized electricity grids (via solar, wind, hydroelectric, and nuclear power) offer significantly reduced carbon emissions related to energy production than those running on non-renewable energies (gas, oil, and coal). The total emissions of each collider project are then evaluated as the product of the total amount of energy consumed and the local carbon intensity for its production. Increasing the sustainability of ILC and CLIC has been a significant point of emphasis since the publications of their technical design report~\cite{ILC_TDR} and conceptual design report~\cite{CLIC_CDR}, respectively. Here we incorporate the latest published updates for the ILC~\cite{ILCInternationalDevelopmentTeam:2022izu} and CLIC~\cite{brunner2022clic} running scenarios which both reduced the operating power requirements significantly.

While total decarbonization of the electric grid by 2040 is a nominal goal, it is not assured. The 2040 projection of carbon intensity based on the stated policies scenario for Japan, China, the European Union, and the USA are roughly 150, 300, 40, and 45~t\gwpnospace/GWh, respectively~\cite{iea}. However, local variations in renewable energy systems implementation is neglected in these estimates; for example, the CERN-based colliders could take advantage of a 50:50 mix of renewable and nuclear energy. Additional mitigation strategies, such as construction of dedicated renewable energy plants, would reduce the carbon impact of operations in other regions. This strategy has been thoroughly investigated by the Green ILC Project~\cite{green_ilc}. A more moderate strategy can be envisioned for \CCCnospace. A 185 MW solar farm could be built with a \$150 million budget~\cite{SEIA2021}, double covering the average power requirement of \CCC\footnote{This estimate considers the power optimizations in Table~\ref{table:power_savings}}, such that excess power could be stored for later use at night~\footnote{The additional cost of selling and purchasing energy through utility companies can be reduced through special contracts and is neglected here}, allowing \CCC to achieve green energy independence. The use of multijunction photovoltaic cell fabrication techniques would improve power conversion efficiency well beyond 30\% that is common in today's cells~\cite{Hunt2013,Young2022}, allowing such a solar farm to be situated on about 5 km$^2$ of land~\cite{Brown2022}.

This estimate relies on energy storage systems supported by regional electricity grids. To better understand the feasibility of scaling all parts of energy production (which may fall under the \CCC project budget) and energy storage infrastructure (which would be funded by the US government, but would nonetheless need investment), we perform a holistic cost estimate. We first note that the energy storage capacity required to supply 150 MW continuously for 12 hours is less than 1\% the expected grid energy storage capacity in 2040~\cite{Denholm2021}, indicating that the US grid should be able to reasonable support operations at this scale using renewable energy. We assume lithium ion batteries~\footnote{Lithium ion batteries are not considered to be viable long term energy storage solutions, instead technologies such as flow batteries and systems based on mechanical potential energy are favored} are the primary energy storage technology with a lifetime of 1000 cycles, experiencing 300 cycles per year with 10\% of battery cost reclaimed through recycling at a base cost of \$125/kWh and \$100/kWh  in 2040 and 2050, respectively~\cite{NREL2022}. We take the cost of energy production of solar to be \$0.80/W~\cite{Brown2022} while taking that of onshore, fixed-bottom offshore and floating offshore wind turbines to be around \$1.3/W, \$3.25/W and \$5.3/W~\cite{Blewett2021, DOE2022}. An energy production portfolio that provides continuous power for \CCC over a 12 hour day/12 hour night period based on these technologies alone would cost approximately \$1 billion. This estimate is primarily driven by requirements of battery energy storage systems and holds for a variety of energy source mixes. This indicates a similar cost would be associated to a site located near the Pacific or Atlantic coasts, which could leverage floating and fixed-bottom turbines respectively, in the southern USA where solar would be most efficient, or proximate to large wind farms in the Midwest. A more precise cost and feasibility analysis can be performed when a candidate site is defined, as has been done for experiments operating at the South pole, for example~\cite{babinec2023feasibility}. This cost analysis demonstrates that \CCC operations could be supported sustainably within the USA within the next two decades given conservative projections of technological development.

As a point of comparison, the power requirement of FCC would be about 30\% of the output of a large nuclear plant (generating 1.1 GW on average~\cite{a2019-frances}). At about \$8 billion per facility, the cost of renewable energy infrastructure for the FCC would be about \$2.5 billion. 
To obtain an estimate of the carbon impact of operations at future collider facilities that takes mitigation strategies into account, we first note that the carbon intensity of solar, wind, hydroelectric, and nuclear are around 30, 15, 25 and 5~ton~\gwpnospace/GWh, respectively~\cite{electricitymaps}. These estimates have some regional variation due to the differences in supply chains and local infrastructure. For instance, given the lifetime of existing nuclear plants of about 30 years, replacement or construction of entirely new facilities will be required and it might effect the overall carbon intensity. While the ultimate energy production portfolio will be different for facilities constructed in different regions, we take a common estimate of 20~t\gwpnospace/GWh for all collider facilities in this analysis. We find this to be a reasonable estimate given that any facility can propose mitigation strategies to decouple their carbon impact from the regional average. It also reflects the expectation that clean energy infrastructure supply chains will improve over the next 20 years.


\section{Analysis of total carbon footprint} 
\label{sec:results}
A straightforward calculation of total energy consumption is possible using the information summarized in Table~\ref{tab:running_scenarios}, which includes estimates of the site power $P$ during collision mode, the annual collision time $T_{\mathrm{collisions}}$ and the total running time in years $T_{\mathrm{run}}$ for each center-of-mass energy $\sqrt{s}$ considered. We take into account the time spent with the beam operating at full rf and cooling power outside of data-taking mode, for example for machine development, as an additional week for every 6 weeks of data-taking (i.e. +17\%), represented as $T_\mathrm{development}$. We take the site power requirement for the remaining period in a calendar year to be 30\% of the site power requirement during data-taking (denoted by $\kappa_\mathrm{down}$). This value is a conservative upper estimate, since without rf power and associated heat load, any accelerator can be kept cold with a small fraction of power to the cryogenics system.

\begin{table}[!ht]
    \centering
    \caption{For each of the Higgs factory projects considered in the 1st row, the center-of-mass energies (2nd row), AC site power (3rd row), annual collision time (4th row), total running time\footnote{The nominal run schedule reflects nominal data-taking conditions, which ignore other run periods such as luminosity ramp-up.} (5th row), instantaneous luminosity per interaction point (6th row) and target integrated luminosity (7th row) at each center-of-mass energy  are given. The numerical values were taken from the references mentioned in the table in conjunction with \cite{Roser:2022sht}. For CEPC the new baseline scenario with 50 MW of synchrotron radiation power per beam is used. We consider both the baseline and the power optimizations of Table~\ref{table:power_savings} (in brackets) for \CCC power requirements.} 
    \label{tab:running_scenarios}
     \resizebox{\columnwidth}{!}{%
    \begin{tabular}{c|c|c c|c c|c c c c|c c c c c c}
    \hline \hline
          Higgs factory & CLIC \cite{brunner2022clic}  & \multicolumn{2}{c|}{ILC \cite{ILCInternationalDevelopmentTeam:2022izu}} & \multicolumn{2}{c|}{\CCC \cite{nanni2022c}} & \multicolumn{4}{c|}{CEPC \cite{cheng2022physics},\cite{cepcacceleratorstudygroup2022snowmass2021}} & \multicolumn{6}{c}{FCC \cite{fcc_snowmass},\cite{Benedikt:2651299}, \cite{fcceeweek_slides}} \\
        $\sqrt{s}$ (GeV) & 380 & 250 & 500 & 250 & 550 &  91.2 & 160 & 240 & 360 & \multicolumn{2}{c}{88,91,94} & 157,163  & 240 & 340-350 & 365 \\ \hline
        $P$ (MW) & 110 & 111 & 173 & 150 (87) & 175 (96) & 283 & 300 & 340 & 430 & \multicolumn{2}{c}{222} & 247  &  273 & \multicolumn{2}{c}{357} \\ 
        $T_{\mathrm{collisions}}$ ($10^{7}$ s/year) & 1.20 & \multicolumn{2}{c|}{1.60} & \multicolumn{2}{c|}{1.60} & \multicolumn{4}{c|}{1.30} & \multicolumn{6}{c}{1.08} \\ 
        $T_{\mathrm{run}}$ (years) & 8 & 11 & 9 & 10 & 10 & 2 & 1 & 10 & 5 & 2 & 2 & 2 &  3 & 1 & 4 \\ 
        $\mathcal{L}_{\mathrm{inst}}/\mathrm{IP} \ (\cdot 10^{34} \ \mathrm{cm}^{-2} \ \mathrm{s}^{-1}  )$ & 2.3 & 1.35 & 1.8 & 1.3 & 2.4 & 191.7 & 26.6 &  8.3 & 0.83 & 115 & 230 & 28 & 8.5 & 0.95 & 1.55 \\ 
        $\mathcal{L}_{\mathrm{int}} \ ( \mathrm{ab}^{-1}  )$ & 1.5 & 2 & 4 & 2 & 4 & 100 & 6 &  20 & 1 & 50 & 100 &  10 & 5 & 0.2 & 1.5 \\ \hline \hline
    \end{tabular}
    }
\end{table}

Using these values, we calculate the annual energy consumed as:

\begin{equation}
    E_{\mathrm{annual}} = P\left[\kappa_\mathrm{down}\cdot T_\mathrm{year}+(1-\kappa_\mathrm{down})(T_\mathrm{collisions} + T_\mathrm{development})\right]
\end{equation}

\noindent and the total energy consumption obtained by summing over all $\sqrt{s}$ run configurations is given by

\begin{equation}
    E_{\mathrm{total}}=\sum_{r\,\in\,\mathrm{runs}}{{E(r)}_{\mathrm{annual}}} \cdot T_{\mathrm{run}}(r)
\end{equation}

For the circular collider projects, the FCC and the CEPC, we consider separately the cumulative energy consumption of the Higgs physics runs (i.e. $\sqrt{s}>240$ GeV) for a focused comparison on the basis of Higgs physics reach argued in Section~\ref{sec:physics_reach}, but additionally include the contribution of $Z$-pole and $WW$-threshold runs which impact the climate nevertheless.

The inclusion of those additional runs enriches the overall physics program of the colliders in a way not reflected in the framework defined in Section~\ref{sec:physics}. The main purpose of the proposed colliders under consideration here is to serve as Higgs factories, and thus we maintain the importance of assessing their physics reach through the projected precision for Higgs observables. Furthermore, as we will demonstrate later, the inclusion of those additional runs does not significantly alter the GWP of the circular machines, which is dominated by construction.

It is worth noting that the FCC-ee tunnel is planned to be reused to host a high energy hadron collider, while a high energy machine following \CCC requires additional construction. However, such a hadron collider requires new structures, notably a dedicated superconducting magnet system, along with the disposal of the $\ee$ beam-line that will have been exposed to high levels of radiation. A full life cycle assessment including the carbon impact of the accelerator structures and end of life plan is required to quantify the relative advantage of reusing the tunnel and is beyond the scope of this work.

\begin{figure}[h!]

\subfloat[\label{fig:energy}]{%
  \includegraphics[scale=0.26]{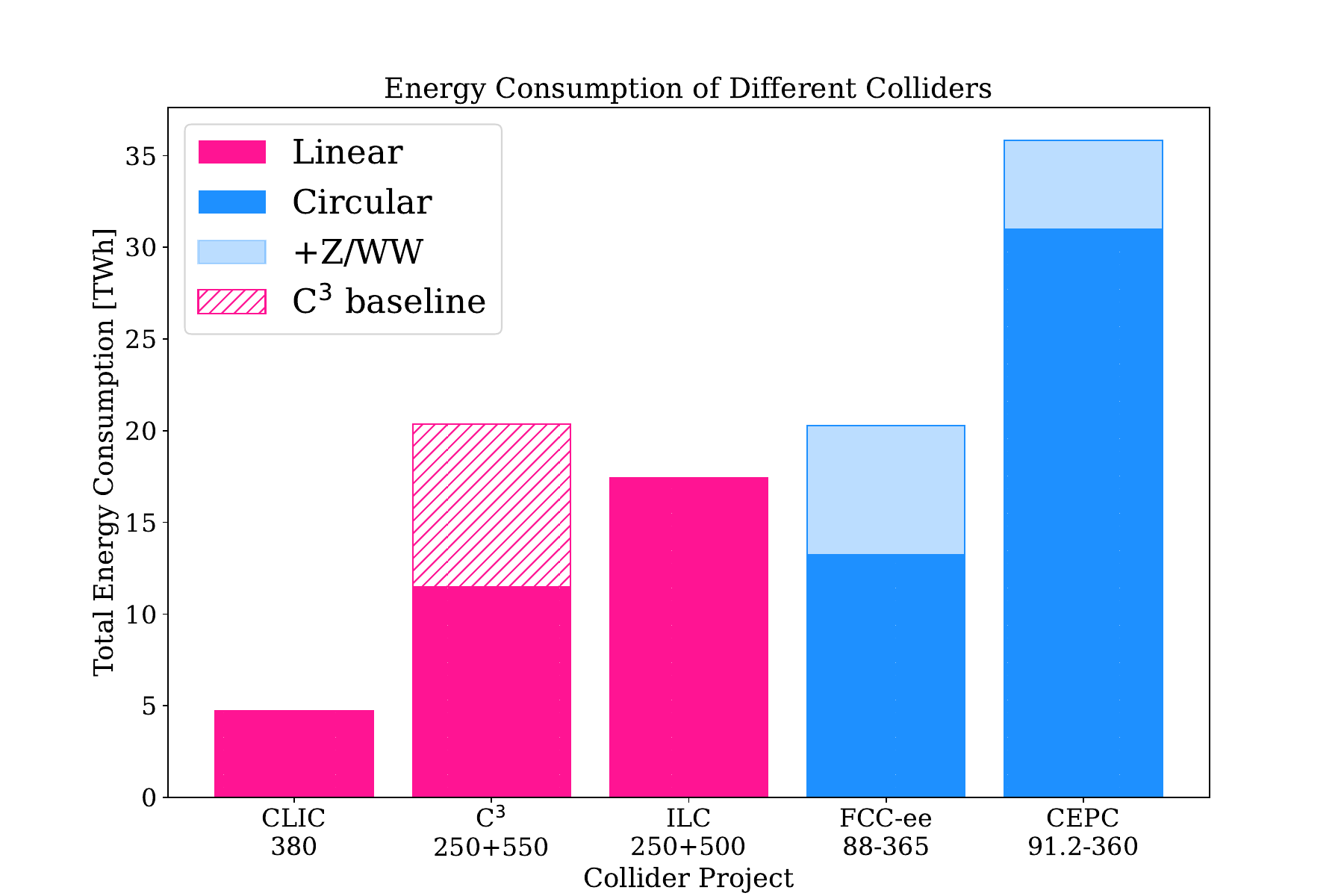}%
}\hfill
\subfloat[\label{fig:weighted_energy}]{%
  \includegraphics[scale=0.26]{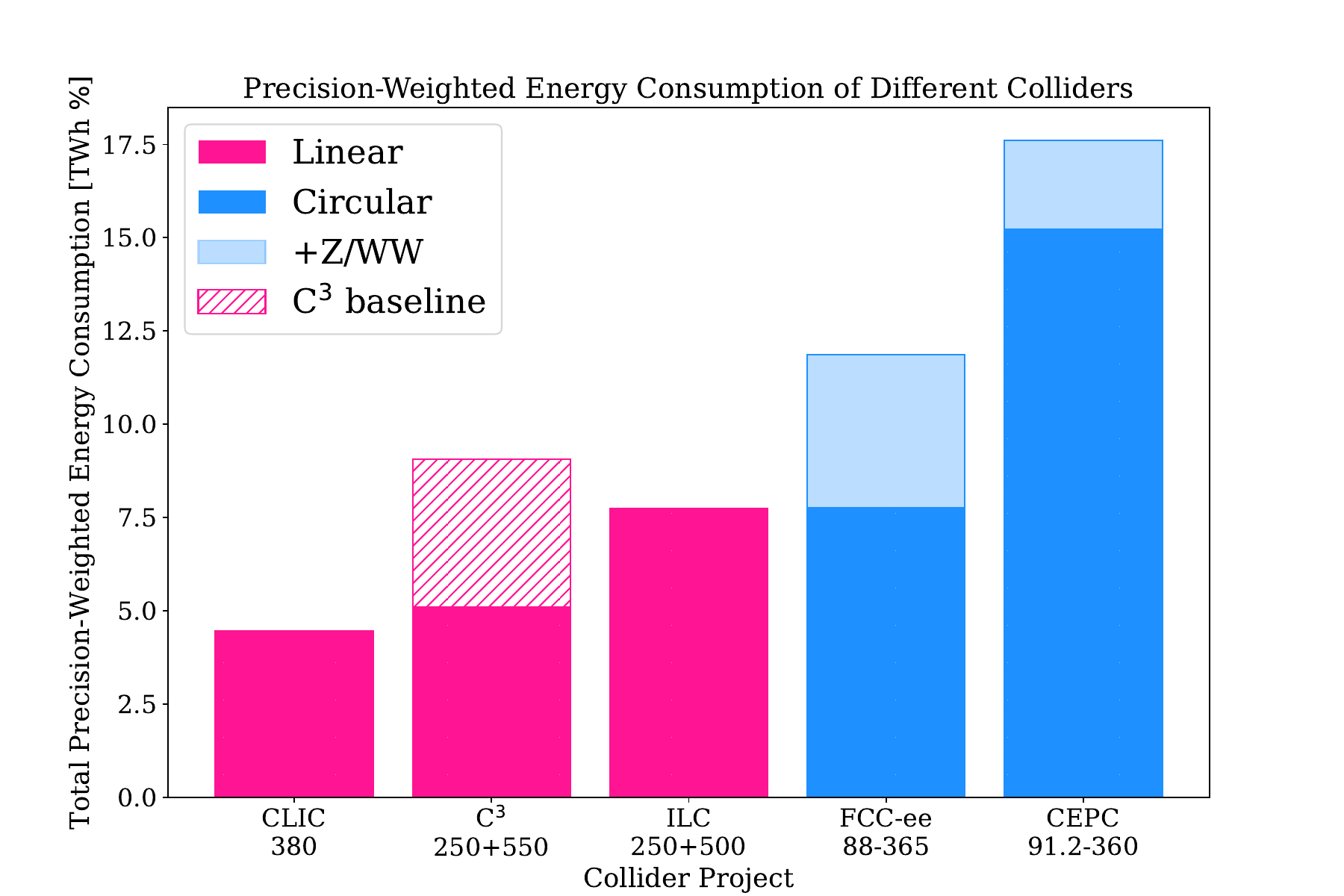}%
}
\caption{Total energy consumption for all collider concepts, both (a) unweighted and (b) weighted with respect to the average coupling precision for each collider. We note that the hashed pink component represents the additional costs of operating \CCC without power optimisations, while light blue regions account for additional run modes targeting $Z$ and $WW$ production.}
\label{fig:total_energy_consumption_unweighted_weighted}
\end{figure}

\begin{figure}[h!]

\subfloat[\label{fig:operations}]{%
  \includegraphics[scale=0.26]{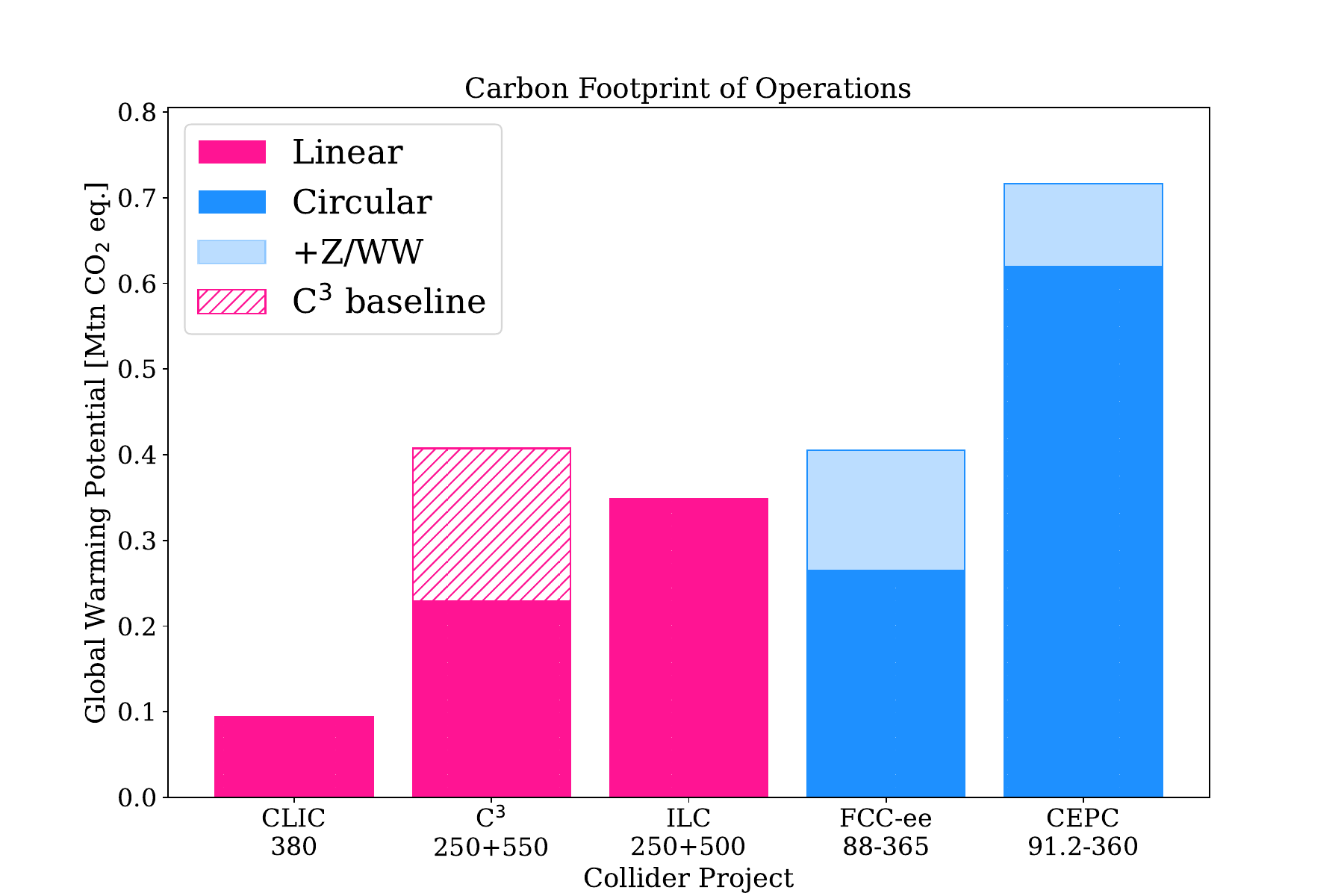}%
}\hfill
\subfloat[\label{fig:construction}]{%
  \includegraphics[scale=0.26]{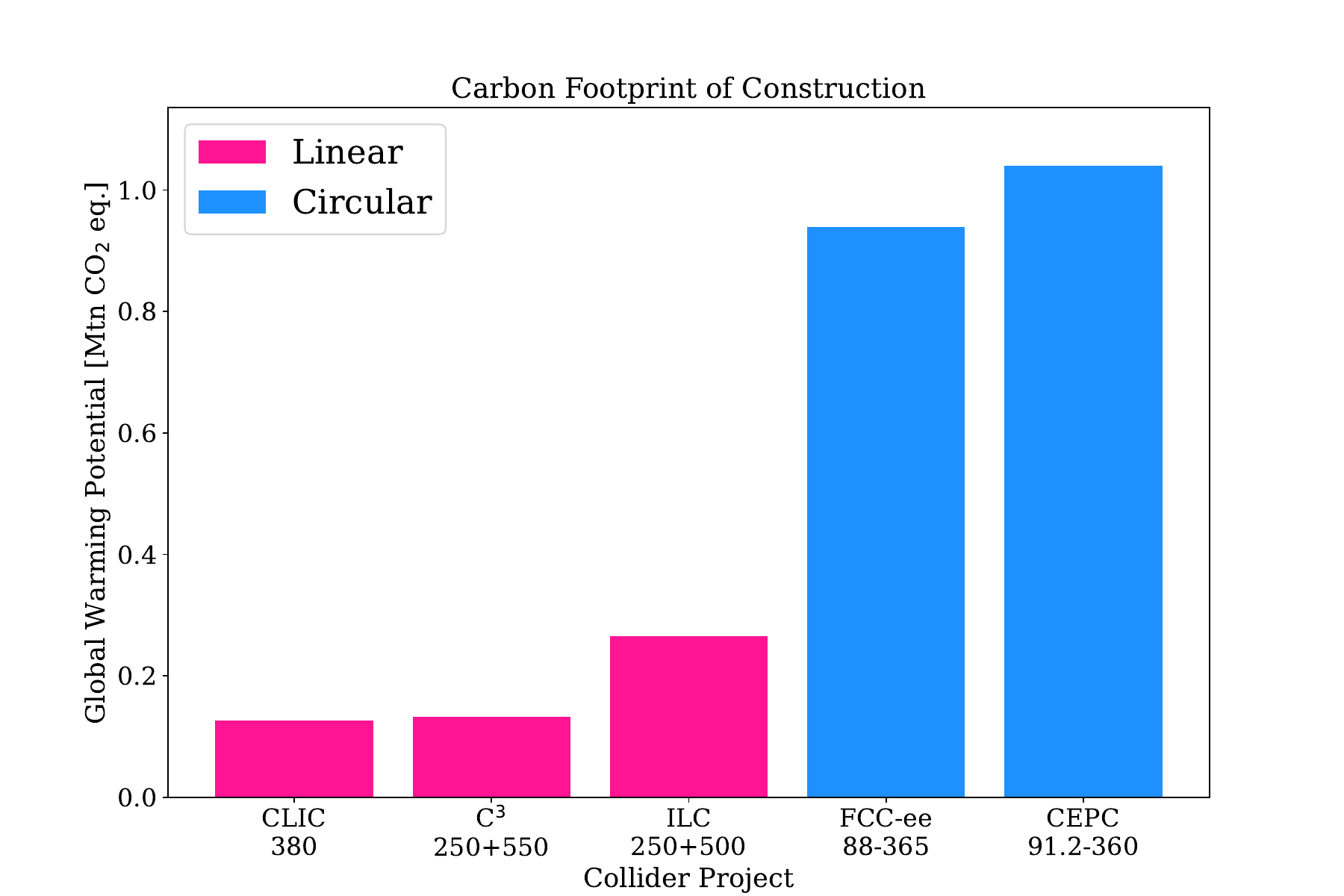}%
}
\caption{Global warming potential from (a) operations and (b) construction of all collider concepts. We note that the hashed pink component represents the additional costs of operating \CCC without power optimisations, while light blue regions account for additional run modes targeting $Z$ and $WW$ production.}
\label{fig:C02_operations_and_construction}
\end{figure}

Figure~\ref{fig:energy} shows the energy consumption for the considered collider projects. The least energy is consumed by CLIC, driven by the lowest planned run time at low energies and its marginally lower power consumption compared to \CCC and the ILC, which are comparable. The energy consumption of the CEPC is large compared to the FCC because the CEPC plans to collect four times the integrated luminosity at 240 GeV with an associated tripling of the total run duration. 

Figure~\ref{fig:weighted_energy} shows the precision-weighted energy consumption for the considered collider projects, estimated by multiplying the energy consumption of Figure~\ref{fig:energy} with the average relative precision in the last row of Table~\ref{tab:higgs_couplings}. The lowest run time for CLIC is now compensated by the reduced relative precision, in comparison to \CCC and ILC, leading to overall closer precision-weighted energy consumption. Similarly, the large proposed run time for the CEPC is now taken into account in conjunction with the improved precision reach, yielding a total weighted energy consumption closer to the FCC.

Figure~\ref{fig:operations} shows the associated GWP of the total energy required for operation, obtained by our multiplying the total energy consumption by the respective carbon intensity. The relative performance among the facilities in terms of GWP is identical to their relative performance in total energy consumption, due to the common carbon intensity of 20 ton~\gwpnospace/GWh taken for all facilities.

\begin{figure}[h!]

\subfloat[\label{fig:total}]{%
  \includegraphics[scale=0.26]{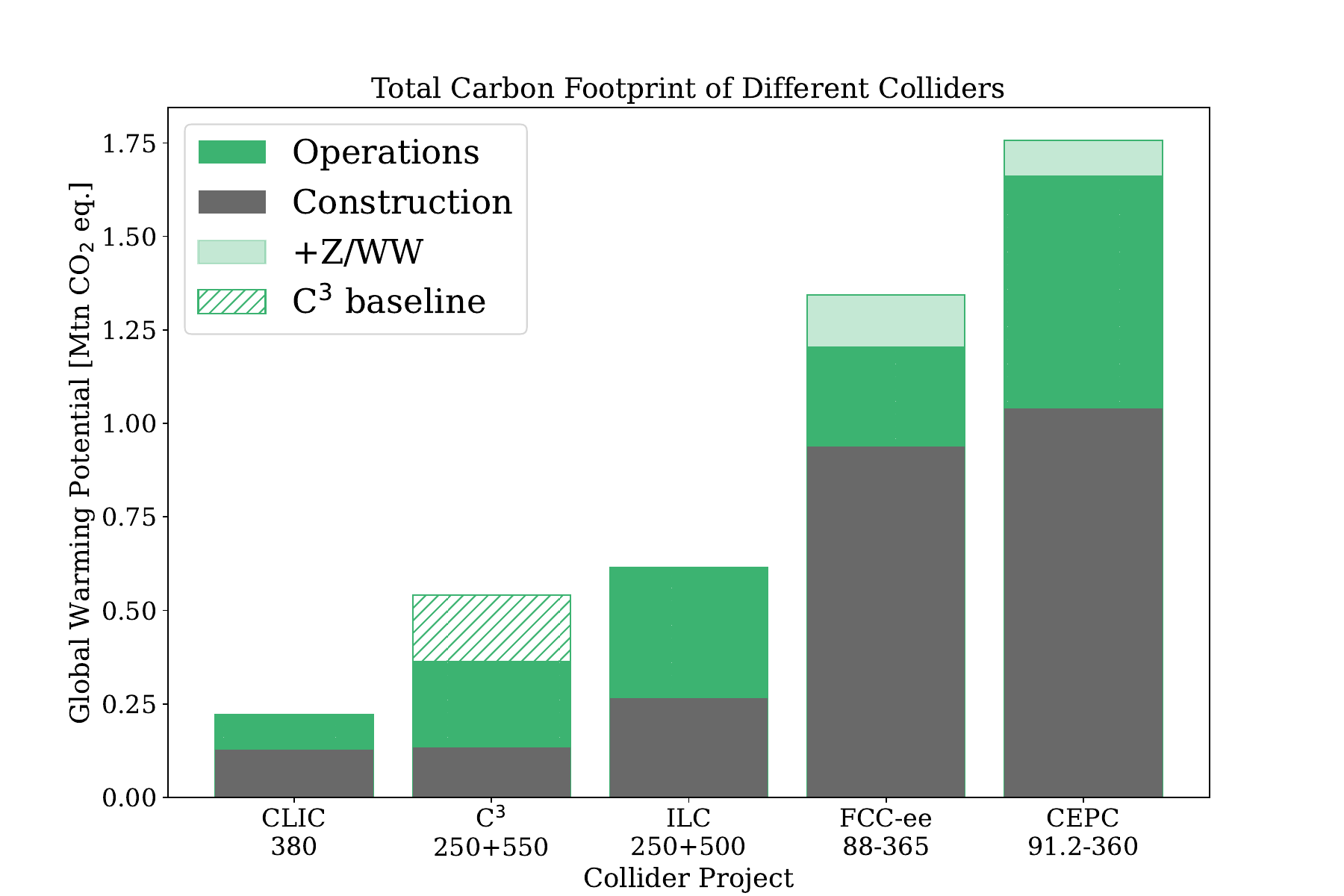}%
}\hfill
\subfloat[\label{fig:weighted_total}]{%
  \includegraphics[scale=0.26]{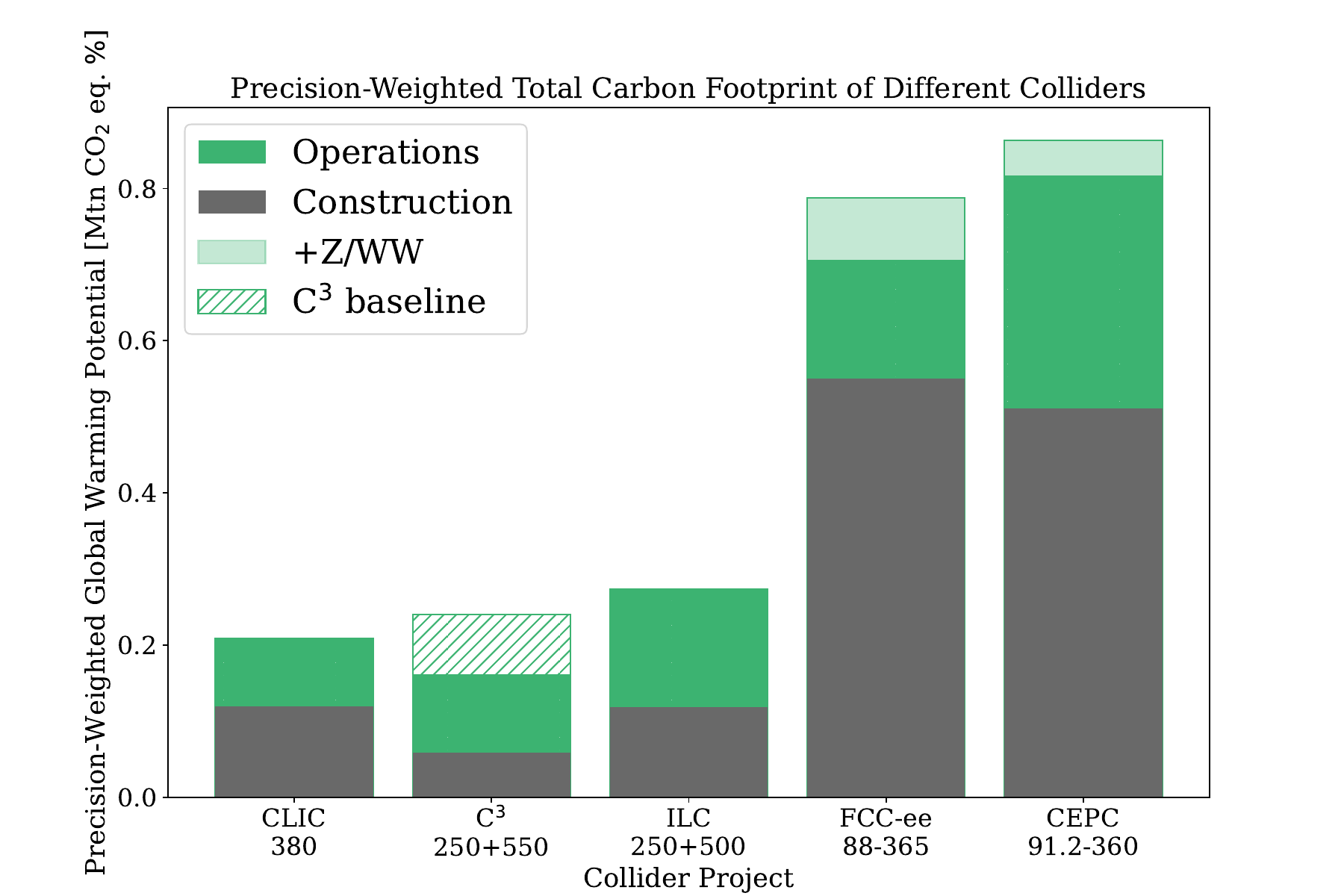}%
}
\caption{Total global warming potential from construction and operations for all collider concepts, both (a) unweighted and (b) weighted with respect to the average coupling precision for each collider. We note that the hashed pink component represents the additional costs of operating \CCC without power optimizations, while light blue regions account for additional run modes targeting $Z$ and $WW$ production.}
\label{fig:total_unweighted_weighted}
\end{figure}

Figure~\ref{fig:construction} shows the GWP due to construction of accelerator facilities. The carbon footprint is very similar among the linear and circular colliders, which is driven primarily by the total length of the accelerator. Figure~\ref{fig:total} shows the total GWP from construction and operations. CLIC is the most environmentally friendly option, owing to its lead performance in operations emissions as well as its small footprint. The total GWP of \CCC and ILC are driven by operations while that of CLIC, the FCC, and the CEPC are almost entirely driven by construction emissions. Possible reductions in the construction component could be achieved by using concrete with lower cement content than CEM1 C40 considered in this analysis. Such cases would still leave the FCC GWP dominated by construction processes.

Finally, Figure~\ref{fig:weighted_total} shows the total precision-weighted GWP from construction and operations, estimated in the same way as the precision-weighted energy consumption in Figure~\ref{fig:weighted_energy}. Given the overall similar GWP for CLIC and \CCC and the  superior precision reach of \CCC at higher energies, compared to CLIC, \CCC appears to be the most environmentally friendly option, when the precision-weighted total carbon footprint is accounted for.

\section{Conclusions}
We present the first analysis of the environmental impact of the newly proposed \CCC collider and a comparison with the other proposed facilities in terms of physics reach, energy needs and carbon footprint for both construction and operations.

The physics reach of the proposed linear and circular $e^+e^-$ colliders has been studied extensively in the context of the US Snowmass and European Strategy processes. We focus on the precision of Higgs boson coupling measurements  achievable at \CCCnospace, CLIC,  the ILC, the FCC, and the CEPC. We point out that in terms of physics reach, all the proposed machines are generally similar, although linear colliders can operate at higher collision energies, enabling access to additional measurements of the Higgs boson's properties. Moreover, the use of polarization at linear facilities effectively compensates for the lower luminosity. 

On this basis, the global warming potential of these facilities is compared in terms of absolute environmental impact and in terms of environmental impact per unit of physics output obtained by a weighted average of expected precision on Higgs coupling measurements. The operations emissions of \CCC could be improved through beam parameter optimization leading to 63 (79) MW power reduction compared to the nominal 150 (175) MW in the 250 (550) GeV running mode. Mitigation strategies using dedicated renewable energy facilities can reduce the carbon intensity of energy production to 20 ton~\gwpnospace/GWh. We find that global warming potential is driven by construction rather than by operations beyond 2040. The compact nature of linear collider facilities reduces the total volume of construction materials and opens up the option for a surface site to simplify the construction process. We conclude that linear colliders and \CCC in particular have great potential for an environmentally sustainable path forward for high energy collider facilities.

\newpage

\section{Acknowledgements}
The authors express their gratitude to Dan Akerib, Tom Shutt, Sridhara Dasu, Patrick Meade, Nigel Lockyer, Sarah Carson and Jim Brau for their insightful discussions, which have significantly contributed to this work. The authors also extend their appreciation to Michael Peskin and Steinar Stapnes for providing feedback on the manuscript.
The work of the authors is supported by the US Department of Energy under contract DE–AC02–76SF00515.

\addcontentsline{toc}{section}{Bibliography}

\bibliographystyle{atlasnote}
\bibliography{sample.bib}

\end{document}